\documentclass[aps,nofootinbib,showpacs]{revtex4}
\usepackage{graphicx}
\usepackage{amsfonts}
\usepackage{amsmath}

\usepackage{color}

\begin{document}

\title{Pair Production of Scalar Dyons in Kerr-Newman Black Holes}

\author{Chiang-Mei Chen} \email{cmchen@phy.ncu.edu.tw}
\affiliation{Department of Physics, National Central University, Chungli 32001, Taiwan}
\affiliation{Center for Mathematics and Theoretical Physics, National Central University, Chungli 32054, Taiwan}

\author{Sang Pyo Kim} \email{sangkim@kunsan.ac.kr}
\affiliation{Department of Physics, Kunsan National University, Kunsan 54150, Korea}

\author{Jia-Rui Sun} \email{sunjiarui@sysu.edu.cn}
\affiliation{School of Physics and Astronomy, Sun Yat-Sen University, Guangzhou 510275, China}

\author{Fu-Yi Tang} \email{foue.tang@gmail.com}
\affiliation{Department of Physics, National Central University, Chungli 32001, Taiwan}

\date{\today}

\begin{abstract}
We study the spontaneous pair production of scalar dyons in the near extremal dyonic Kerr-Newman (KN) black hole, which contains a warped AdS$_3$ structure in the near horizon region. The leading term contribution of the pair production rate and the absorption cross section ratio are also calculated using the Hamilton-Jacobi approach and the thermal interpretation is given. In addition, the holographic dual conformal field theories (CFTs) descriptions of the pair production rate and absorption cross section ratios are analyzed both in the $J$-, $Q$- and $P$-pictures respectively based on the threefold dyonic KN/CFTs dualities.
\end{abstract}


\maketitle
\section{Introduction}
The spontaneous pair production is associated with the separation of the virtual particle pairs, which can be caused by an external electromagnetic force or by the tunnelling through a causal boundary. These effects are related with two independent quantum processes: the Schwinger mechanism~\cite{Schwinger:1951nm} and the Hawking radiation~\cite{Parikh:1999mf}. A charged black hole can provide a suitable arena to address both processes. Moreover, it is shown that the leading contribution of the Schwinger effect comes from the dominating electromagnetic field in the near horizon region of the charged black hole and the pair production rate is shown to match with that obtained in the whole black hole background~\cite{Chen:2012zn, Chen:2014yfa, Chen:2015fkx, Chen:2016caa}. Besides, for the near extremal charged black holes, the symmetry of the near horizon geometry is enhanced, e.g. AdS$_2 \times S^2$ for the Reissner-Nordstr\"om (RN), and a warped AdS$_3 = $ AdS$_2 \times S^1$ for the Kerr-Newman (KN) black hole, which leads to a holographic dual conformal field theory (CFT) description of the black holes~\cite{Guica:2008mu, Chen:2009ht, Chen:2010bsa, Chen:2010as, Chen:2010yu, Chen:2011gz, Chen:2010ywa, Chen:2012np, Compere:2012jk}.

Recently, the pair production of scalar field~\cite{Chen:2012zn} and spinor field~\cite{Chen:2014yfa} in the near horizon region of RN black holes, as well as pair production of scalar field in near extremal KN black holes~\cite{Chen:2016caa} have been systematically analyzed. An interesting finding was that the violation of the Breitenlohner-Freedman (BF) bound in the AdS$_2$ or the warped AdS$_3$ geometry is equivalent to the existence condition for the pair production, and also guarantees the cosmic censorship conjecture during the pair production process. This condition, viewed in the AdS/CFT correspondence, results in the ``complex'' conformal dimensions of operators in the dual CFT, which indicates the dual (super)gravity solution is unstable~\cite{Distler:1998a}. More interestingly, the pair production from the near extremal RN or KN black holes exhibit a thermal interpretation~\cite{Kim2015a, Kim2015b, Chen:2016caa}, in which the production rate depends on an effective temperature coming from a combination of the Unruh temperature caused by the acceleration driven by the electromagnetic force and the AdS curvature~\cite{Cai2014}. Moreover, the particle mass is modified by the AdS curvature and the quantum number of the compact space, which reduces the Schwinger effect in AdS$_2$~\cite{Cai2014, Chen:2016caa, Kim2015a}. This effect is analogous to the transverse momentum of particle spectrum produced by an electric field in quantum electrodynamics (QED) in the Minkowski spacetime. Indeed, the pair production rates in both the near extremal RN and KN black holes can be understood as a ``combination'' of two contributions of Schwinger mechanism: one is the effect from the AdS$_2$ space~\cite{Cai2014} determined by the effective temperature, another one is from the Rindler space associated with the Hawking temperature of black holes~\cite{Kim2015a, Chen:2016caa, Kim2015b}. The holographic dual of the non-extremal KN black hole has been shown to have the twofold dual CFT descriptions, called the $J$-picture and the $Q$-picture~\cite{Chen:2010ywa, Chen:2012np}. The $J$-picture is shown to be associated with black hole angular momentum and can be interpreted by the Kerr/CFT~\cite{Guica:2008mu, Compere:2012jk}, and the $Q$-picture is shown to be associated with the gauge potential, which provides the $U(1)$ fiber and can be interpreted by the RN/CFT~\cite{Chen:2009ht, Chen:2010bsa, Chen:2010as, Chen:2010yu, Chen:2011gz}. The scalar pair production of the near extremal KN black hole has been shown to inherit the same twofold dual CFT descriptions by comparing of the absorption cross section ratios of scalar fields in the near extremal KN black hole with those of the scalar operators in the dual CFTs~\cite{Chen:2016caa}.

In the present paper, we extend our previous study~\cite{Chen:2016caa} to dyonic KN black holes, which contain the magnetic charges in addition to electric charges~\cite{Chen:2010yu, Chen:2012ps, Chen:2012pt}. To study the effect of magnetic charges on the Schwinger effect, the probe scalar field is required to carry both the electric and magnetic charges, which can be viewed as a mixture of electric charged particles and magnetic charged monopoles. Thanks to the warped AdS$_3$ structure appeared in the near horizon region of the near extremal dyonic KN black hole (a slightly difference is that the AdS curvature radius is enlarged due to the magnetic charges which implies the curvature is decreased), the information of its dual CFT descriptions can be easily obtained. We thus study the spontaneous pair production of the charged scalar dyons in the near horizon near extremal dyonic KN black hole following the method in~\cite{Chen:2016caa} without considering the back reactions. By investigating the thermal interpretation of the Schwinger pair production from the near extremal dyonic KN black hole, we show that, similar to the electric charged case~\cite{Chen:2016caa}, the production rate can also be divided into two contributions in AdS and Rindler spaces, and each term is modified by the magnetic charge contributions. Again, to further understand the physical origin of the leading term of the production rate, we apply the Hamilton-Jacobi action approach and compute the phase integral in the complex plane to obtain the instanton action. There are three simple poles located at inner/outer horizons, and asymptotic boundary (still in the near horizon region of the original KN black hole). Actually the associated residues correspond to the three critical parameters appearing in the solution of the radial equation. In particular, the Boltzmann factor of the Schwinger effect is the consequence of the intertwining between the residue contributions at the inner/outer horizons and at the asymptotic boundary. So it likely also includes the quantum tunnelling process from the inner to outer horizons of the dyonic KN black holes. Finally, we study the holographic dual CFTs descriptions of the Schwinger pair production in the dyonic KN black hole in terms of the KN/CFTs duality and the dyonic RN/CFT duality~\cite{Chen:2010yu, Chen:2010ywa, Chen:2012pt}. In addition to the $J$-picture and $Q$-picture, there is a new magnetic charged $P$-picture in the dual CFT descriptions, which is connected with the $Q$-picture via the electromagnetic duality in four dimensional spacetime. We verify that the absorption cross section ratios of scalar dyons in the near extremal dyonic KN black hole match with those of the scalar operator in the dual CFT, in each of the three pictures respectively, which provide a clear holographic description of the Schwinger pair production in the bulk.

The rest of the paper is organized as follows. In Sec.\ref{sec2}, a quick review of the dyonic Kerr-Newman black hole and its near horizon near extremal counterpart is given. Then in Sec.\ref{sec3}, an analytical solution of the probe charged (with electric and magnetic charges) scalar field in this background is calculated and the associated pair production rate as well as the absorption cross section ratio are obtained. In Sec.\ref{sec4} and Sec.\ref{sec5}, the thermal interpretation and dual CFTs descriptions are presented, respectively. Finally, the conclusion is drawn in Sec.\ref{sec6}.

\section{Near Horizon of Near Extremal Dyonic Kerr-Newmann Black Holes}\label{sec2}
The most general solution of the four-dimensional Einstein-Maxwell theory
\begin{equation}
S = \frac1{16 \pi} \int d^4x \sqrt{-g} \left( R - F_{[2]}^2 \right),
\end{equation}
is the dyonic Kerr-Newmann black holes carrying four physical parameters, mass $M$, angular momentum $J = M a$, electric charge $Q$ and magnetic charge $P$, as given by
\begin{eqnarray}\label{dyonic KN}
ds^2 &=& - \frac{\Sigma \Delta}{(\hat{r}^2 + a^2)^2 - \Delta a^2 \sin^2\theta} d\hat{t}^2 + \frac{\Sigma}{\Delta} d\hat{r}^2 + \Sigma d\theta^2
\nonumber\\
&& + \frac{(\hat{r}^2 + a^2)^2 - \Delta a^2 \sin^2\theta}{\Sigma} \sin^2\theta \left[ d\hat{\varphi} - \frac{a (2 M \hat{r} - Q^2-P^2)}{(\hat{r}^2 + a^2)^2 - \Delta a^2 \sin^2\theta} d\hat{t} \right]^2,
\\
A_{[1]} &=& \frac{Q \hat r - P a \cos\theta}{\Sigma} \left( d\hat t - a \sin^2\theta d\hat\varphi \right) +P\left( \cos\theta\mp 1\right)d\hat\varphi, \label{A1}
\end{eqnarray}
where
\begin{equation}
\Sigma = \hat r^2 + a^2 \cos^2\theta, \qquad \Delta = \hat r^2 - 2 M \hat r + a^2 + Q^2 + P^2.
\end{equation}
The magnetic monopole induces a string-like singularity causing different choices of the gauge potential: the upper sign is regular in $0 \leq \theta < \pi/2$, and the lower sign in $\pi/2 < \theta \leq \pi$~\cite{Wu:1975es, Semiz:1991kh}. The inner and outer horizons are located at $r_\pm = M \pm \sqrt{M^2 - a^2 - Q^2 - P^2}$, and the extremal limit corresponds to $M^2 = a^2 + Q^2 + P^2$, in which the horizons degenerate to $r_+ = r_- = M = \sqrt{a^2 + Q^2 + P^2} \equiv r_0$. The dual gauge potential, $\bar A_{[1]}$, is defined as (hereafter ${}^\star$ is the Hodge dual)
\begin{equation}
\bar F_{[2]} = d \bar A_{[1]} \quad \textrm{such that} \quad  \bar F_{[2]} = {}^\star F_{[2]}.
\end{equation}
Thus the dual potential of~(\ref{A1}) is
\begin{equation}
\bar A_{[1]} = \frac{P \hat r + Q a \cos\theta}{\Sigma} \left( d\hat t - a \sin^2\theta d\hat\varphi \right) -  Q(\cos\theta\mp 1) d\hat\varphi. \label{bA1}
\end{equation}
The thermodynamical properties of dyonic KN black holes are described by two essential quantities, the Hawking temperature and the Bekenstein-Hawking entropy
\begin{equation}
T_\mathrm{H} = \frac{r_+ - r_-}{4 \pi (r_+^2 + a^2)}, \qquad S_\mathrm{BH} = \pi (r_+^2 + a^2),
\end{equation}
beside, the associated electric and magnetic chemical potentials and the angular velocity are
\begin{equation}
\Phi_\mathrm{H} = - \frac{Q r_+}{r_+^2 + a^2}, \qquad \bar\Phi_\mathrm{H} = - \frac{P r_+}{r_+^2 + a^2}, \qquad \Omega_\mathrm{H} = \frac{a}{r_+^2 + a^2}.
\end{equation}

To derive the near horizon geometry for the near extremal black holes, one first considers the corotating coordinates with the horizon angular velocity $\Omega_0 = a/(r_0^2 + a^2)$ by
\begin{equation}
\hat{\varphi} \to \varphi + \frac{a}{r_0^2 + a^2} \hat{t},
\end{equation}
and then takes the following near horizon and near extremal limit with $\varepsilon \to 0$,
\begin{equation} \label{NHKNlimit}
\hat{r} \to r_0 + \varepsilon r, \qquad \hat{t} \to \frac{r_0^2 + a^2}{\varepsilon} t, \qquad M \to r_0 + \varepsilon^2 \frac{B^2}{2 r_0}.
\end{equation}
to obtain the near horizon solution of the near extremal dyonic KN black hole as
\begin{eqnarray} \label{NHKN}
ds^2 &=& \Gamma(\theta) \left[ -(r^2 - B^2) dt^2 + \frac{dr^2}{r^2 - B^2} + d\theta^2 \right] + \gamma(\theta) (d\varphi + b r dt)^2,
\\
A_{[1]} &=& - \frac{Q (r_0^2 - a^2 \cos^2\theta) - 2 P r_0 a \cos\theta}{\Gamma(\theta)} r dt - \frac{Q r_0 a \sin^2\theta - P (r_0^2 + a^2) \cos\theta \pm P\Gamma(\theta)}{\Gamma(\theta)} d\varphi, \label{A1n}
\\
\bar A_{[1]} &=&  -\frac{P (r_0^2 - a^2 \cos^2\theta) + 2 Q r_0 a \cos\theta}{\Gamma(\theta)} r dt - \frac{P r_0 a \sin^2\theta + Q (r_0^2 + a^2) \cos\theta \mp Q\Gamma(\theta)}{\Gamma(\theta)} d\varphi, \label{bA1n}
\end{eqnarray}
where
\begin{equation}
\Gamma(\theta) = r_0^2 + a^2 \cos^2\theta, \qquad \gamma(\theta) = \frac{(r_0^2 + a^2)^2 \sin^2\theta}{r^2_0 + a^2 \cos^2\theta}, \qquad b = \frac{2 a r_0}{r_0^2 + a^2}.
\end{equation}
The spacetime~(\ref{NHKN}) contains a warped AdS$_3$ geometry, which allows the dual CFTs description for the KN black hole. The thermodynamical quantities of back holes reduce to
\begin{equation}
T_\mathrm{H} = \frac{B}{2 \pi}, \quad S_\mathrm{BH} = \pi (r_0^2 + a^2 + 2 \varepsilon B r_0), \quad \Phi_\mathrm{H} = \frac{Q (Q^2 + P^2) B}{r_0^2 + a^2}, \quad \bar\Phi_\mathrm{H} = \frac{P (Q^2 + P^2) B}{r_0^2 + a^2}, \quad \Omega_\mathrm{H} = - \frac{2 a r_0 B}{r_0^2 + a^2}.
\end{equation}

\section{Solution of the probe scalar field}\label{sec3}
The action for a probe dyonic charged scalar field $\Phi$ with mass $m$, electric charge $q$ and magnetic charge $p$ is
\begin{equation}
S = \int d^4x \sqrt{-g} \left( -\frac12 D_\alpha \Phi^* D^\alpha \Phi - \frac12 m^2 \Phi^* \Phi \right)
\end{equation}
where $D_\alpha \equiv \nabla_\alpha - i q A_\alpha - i p \bar A_\alpha$. The corresponding Klein-Gordon (KG) equation is
\begin{equation}
(\nabla_\alpha - i q A_\alpha - i p \bar A_\alpha) (\nabla^\alpha - i q A^\alpha - i p \bar A^\alpha) \Phi - m^2 \Phi = 0.
\end{equation}
The ansatz for the scalar field is~\cite{Semiz:1991kh}
\begin{equation} \label{ansatz}
\Phi(t, r, \theta, \varphi) = \mathrm{e}^{-i \omega t + i [n \mp (q P - p Q)] \varphi} R(r) S(\theta),
\end{equation}
and the KG equation can be separated to the angular part
\begin{equation}
\frac1{\sin\theta} \partial_\theta (\sin\theta \partial_\theta S) - \left( \frac{[ n (r_0^2 + a^2 \cos^2\theta) + (q Q + p P) r_0 a \sin^2\theta - (q P - p Q) (r_0^2 + a^2) \cos\theta ]^2}{(r_0^2 + a^2)^2 \sin^2\theta} - m^2 a^2 \sin^2\theta - \lambda_l \right) S = 0,
\end{equation}
and the radial part
\begin{equation} \label{EqR}
\partial_r \left[ (r^2 - B^2) \partial_r R \right] + \left( \frac{\left[ \omega (r_0^2 + a^2) - (q Q + p P) (Q^2 + P^2) r + 2 n a r_0 r \right]^2}{(r_0^2 + a^2)^2 (r^2 - B^2)} - m^2 (r_0^2 + a^2) - \lambda_l \right) R = 0,
\end{equation}
where $\lambda_l$ is the separation constant. The radial equation can be interpreted as the equation of motion of the scalar with an effective mass
\begin{equation}
m^2_\mathrm{eff} = m^2 - \frac{[2 n a r_0 - (q Q + p P)(Q^2 + P^2)]^2}{(r_0^2 + a^2)^3} + \frac{\lambda_l}{r_0^2 + a^2},
\end{equation}
propagating in the AdS$_2$ space with radius $L_\mathrm{AdS} = r_0^2 + a^2$. The instability in the AdS$_2$ space occurs when the BF bound is violated, i.e., $m^2_\mathrm{eff}<-\frac{1}{4L^2_\mathrm{AdS}}$, the corresponding relation is
\begin{equation} \label{BFbv}
m^2_\mathrm{eff} < - \frac{1}{4 (r_0^2 + a^2)} \quad \Rightarrow \quad \frac{[2 n a r_0 - (q Q + p P)(Q^2 + P^2)]^2}{(r_0^2 + a^2)^2} - m^2 (r_0^2 + a^2) - \left( \lambda_l + \frac14 \right) > 0.
\end{equation}

The radial flux of scalar field~(\ref{ansatz}) in the spacetime~(\ref{NHKN}) can be expressed as
\begin{eqnarray} \label{fluxD}
D &=& \int d\theta d\varphi i \sqrt{-g} g^{rr} (\Phi D_r \Phi^* - \Phi^* D_r \Phi)
\nonumber\\
&=& i (r_0^2 + a^2) (r^2 - B^2) (R \partial_r R^* - R^* \partial_r R) \mathfrak{S},
\end{eqnarray}
where the contribution from the angular part is symbolically denoted by
\begin{equation} \label{frakS}
\mathfrak{S} = 2 \pi \int d\theta \sin\theta S \, S^*,
\end{equation}
which will contribute the ``same'' factor for the flux at the near horizon region and at the asymptotic region. Therefore, it does not show up in the flux ratios, and consequently, it will not affect the physical quantities such as the mean number of produced pairs and the absorption cross section ratio, etc.

The general solution of the radial equation~(\ref{EqR}) can be found in terms of the hypergeometric functions,
\begin{eqnarray} \label{sol}
R(r) &=& c_1 (r - B)^{\frac{i}2 (\tilde \kappa - \kappa)} (r + B)^{\frac{i}2 (\tilde \kappa + \kappa)} \, F\left( \frac12 + i \tilde \kappa + i \mu, \frac12 + i \tilde \kappa - i \mu; 1 + i \tilde \kappa - i \kappa; \frac12 - \frac{r}{2B} \right)
\nonumber\\
&+& c_2 (r - B)^{- \frac{i}2 (\tilde \kappa - \kappa)} (r + B)^{\frac{i}2 (\tilde \kappa + \kappa)} \, F\left( \frac12 + i \kappa + i \mu, \frac12 + i \kappa - i \mu; 1 - i \tilde \kappa + i \kappa; \frac12 - \frac{r}{2B} \right),
\end{eqnarray}
with three essential parameters
\begin{equation} \label{kappamu}
\tilde \kappa = \frac{\omega}{B}, \qquad \kappa = \frac{(q Q + p P) (Q^2 + P^2) - 2 n a r_0}{r_0^2 + a^2}, \qquad \mu = \sqrt{\kappa^2 - m^2 (r_0^2 + a^2) - \lambda_l - \frac14},
\end{equation}
in which $\mu^2$ is positive due to the BF bound violation in Eq.~(\ref{BFbv}).

According to the results in Refs.~\cite{Chen:2012zn, Chen:2016caa}, the vacuum persistence amplitude and the mean number respectively are
\begin{eqnarray} \label{Bogoliubovalpha}
|\alpha|^2 &=& \frac{\cosh(\pi \kappa - \pi \mu) \cosh(\pi \tilde\kappa + \pi \mu)}{\cosh(\pi \kappa + \pi \mu) \cosh(\pi \tilde\kappa - \pi \mu)},
\\ \label{Bogoliubovbeta}
|\beta|^2 &=& \frac{\sinh(2 \pi \mu) \sinh(\pi \tilde\kappa - \pi \kappa)}{\cosh(\pi \kappa + \pi \mu) \cosh(\pi \tilde\kappa - \pi \mu)},
\end{eqnarray}
and the absorption cross section ratio is
\begin{equation} \label{Nsigma}
\sigma_\mathrm{abs} = \frac{\sinh(2 \pi \mu) \sinh(\pi \tilde \kappa - \pi \kappa)}{\cosh(\pi \kappa - \pi \mu) \cosh(\pi \tilde \kappa + \pi \mu)},
\end{equation}
which can be further rewritten in the following form
\begin{equation} \label{absorption}
\sigma_\mathrm{abs} = \frac{\sinh(2 \pi \mu)}{\pi^2} \sinh(\pi \tilde\kappa - \pi \kappa) \left| \Gamma\left( \frac12 + i (\mu - \kappa) \right) \right|^2 \left| \Gamma\left( \frac12 + i (\mu + \tilde\kappa) \right) \right|^2.
\end{equation}
Again, the pair production rate $|\beta|^2$ and the absorption cross section ratio $\sigma_\mathrm{abs}$ are connected by the relation $|\beta|^2(\mu\to -\mu)\to -\sigma_\mathrm{abs}$.

\section{Thermal Interpretation}\label{sec4}
Following our previous studies~\cite{Chen:2012zn, Chen:2016caa}, the mean number of produced pairs~(\ref{Bogoliubovbeta}) can be reexpressed as
\begin{equation} \label{meannumberpairs}
\mathcal{N} = |\beta|^2 = \left( \frac{\mathrm{e}^{- 2 \pi \kappa + 2 \pi \mu} - \mathrm{e}^{- 2 \pi \kappa - 2 \pi \mu}}{1 + \mathrm{e}^{- 2 \pi \kappa - 2 \pi \mu}} \right) \left( \frac{1 - \mathrm{e}^{- 2 \pi \tilde\kappa + 2 \pi \kappa}}{1 + \mathrm{e}^{- 2 \pi \tilde\kappa + 2 \pi \mu}} \right).
\end{equation}
Note that the mean number~(\ref{meannumberpairs}) has a similar form except for different quantum numbers as that of charged scalars in a near-extremal RN~\cite{Chen:2012zn} and KN~\cite{Chen:2016caa} since the near-horizon geometry is an ${\rm AdS}_2 \times S^2$ for the near-extremal RN black hole while it is a warped AdS$_3$ for the near-extremal KN black hole. Following Refs.~\cite{Kim2015a, Kim2015b}, we introduce an effective temperature and its associated one
\begin{equation} \label{effectivetemperature}
T_\mathrm{KN} = \frac{\bar{m}}{2 \pi \kappa - 2 \pi \mu} = T_U + \sqrt{T_U^2 + \frac{\mathcal{R}}{8 \pi^2}}, \qquad \bar{T}_\mathrm{KN} = \frac{\bar{m}}{2 \pi \kappa + 2 \pi \mu} = T_U - \sqrt{T_U^2 + \frac{\mathcal{R}}{8 \pi^2}},
\end{equation}
where the effective mass $\bar{m}$ is
\begin{equation} \label{barm}
\bar{m} = \sqrt{m^2 - \frac{\lambda + 1/4}{2} \mathcal{R}},
\end{equation}
and the corresponding Unruh temperature $T_U$ and AdS curvature $\mathcal{R}$ are
\begin{equation} \label{unruh}
T_U = \frac{\kappa}{2 \pi \bar{m}(r_0^2 + a^2)} = \frac{(q Q + p P)(Q^2 + P^2) - 2 n a r_0}{2 \pi \bar{m}(r_0^2 + a^2)^2}, \qquad \mathcal{R} = - \frac{2}{r_0^2 + a^2}.
\end{equation}
Therefore we have
\begin{equation}
2 \pi \kappa - 2 \pi \mu = \frac{\bar{m}}{T_\mathrm{KN}}, \qquad  2 \pi \kappa + 2 \pi \mu = \frac{\bar{m}}{\bar{T}_\mathrm{KN}}, \qquad 2 \pi \tilde\kappa = \frac{\omega}{T_\mathrm{H}}, \qquad 2 \pi \kappa = \frac{q \Phi_\mathrm{H} + p \bar\Phi_\mathrm{H} + n \Omega_H}{T_\mathrm{H}},
\end{equation}
and the mean number~(\ref{meannumberpairs}) can be expressed as
\begin{equation} \label{thermal}
\mathcal{N} = \mathrm{e}^{\frac{\bar{m}}{T_\mathrm{KN}}} \times \left( \frac{\mathrm{e}^{-\frac{\bar{m}}{T_\mathrm{KN}}} - \mathrm{e}^{-\frac{\bar{m}}{\bar{T}_\mathrm{KN}}}}{1 + \mathrm{e}^{-\frac{\bar{m}}{\bar{T}_\mathrm{KN}}}} \right) \times \left\{ \frac{\mathrm{e}^{-\frac{\bar{m}}{T_\mathrm{KN}}} \left( 1 - \mathrm{e}^{-\frac{\omega - q \Phi_\mathrm{H} -p \bar\Phi_\mathrm{H} - n \Omega_\mathrm{H}}{T_\mathrm{H}}} \right)}{1 + \mathrm{e}^{-\frac{\omega - q \Phi_\mathrm{H} - p \bar\Phi_\mathrm{H} - n \Omega_\mathrm{H}}{T_\mathrm{H}}} \mathrm{e}^{- \frac{\bar{m}}{T_\mathrm{KN}}}} \right\}.
\end{equation}
The physical interpretation of each term of Eq.~(\ref{thermal}) is that the first parenthesis is the Schwinger effect with the effective temperature $T_\mathrm{KN}$ in ${\rm AdS}_2$~\cite{Cai2014} and the second parenthesis is the Schwinger effect in the Rindler space~\cite{Gabriel2000}, in which the Unruh temperature is given by the Hawking temperature and the charges have the chemical potentials of $\Phi_\mathrm{H}$, $\bar\Phi_\mathrm{H}$, and $\Omega_\mathrm{H}$, while the effective temperature for the Schwinger effect due to the electric field on the horizon is determined by $T_\mathrm{KN}$.

The mean number of produced pairs above and the absorption cross section ratio in the previous section have been obtained using the exact solution in the near horizon geometry of an extremal or near extremal KN black hole. Below, by applying the phase-integral formula, we derive the instanton actions from the Hamilton-Jacobi action for the field equation, which lead to the mean number. This method allows one to understand the physical origin of each term as a consequence of simple poles in the complex plane of space~\cite{Kim2007} and further also connects the interpretation to other physical systems involving the Schwinger effect in curved spacetimes~\cite{Kim2013}.

\begin{figure}
\centering
\includegraphics[width=4in]{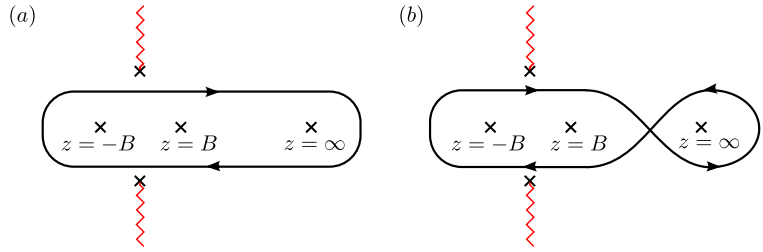}
\caption{The contour integral for the Schwinger effect in AdS space.}
\label{fig1}
\end{figure}

\begin{figure}
\centering
\includegraphics[width=4in]{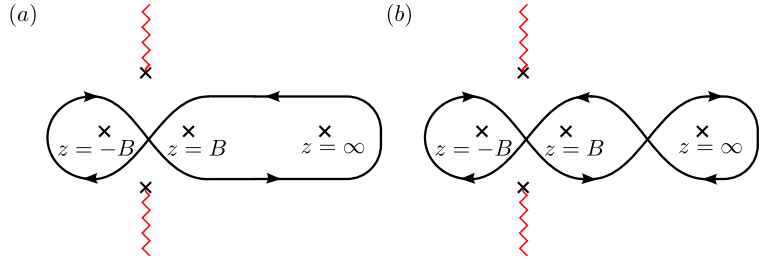}
\caption{The contour integral for the Schwinger effect in Rindler space.}
\label{fig2}
\end{figure}

In order to understand the leading term of the men number, we use the Hamilton-Jacobi approach to pair production via quantum tunneling. In the phase-integral method the Hamilton-Jacobi approximation $R(r) = e^{i S(r)}$ to Eq.~(\ref{EqR}) leading to
\begin{equation} \label{HJeq}
S(r) = \oint \frac{dr}{r^2 - B^2} \sqrt{(\omega - \kappa r)^2 - \bar m^2 (r_0^2 + a^2) (r^2 - B^2)},
\end{equation}
where the effective mass is given by~(\ref{barm})
\begin{equation}
\bar m = m \sqrt{1 + \frac{\lambda_l + 1/4}{m^2 (r_0^2 + a^2)}}.
\end{equation}
There are three simple poles located at $z = \pm B$ and $z = \infty$ and their residue contributions in the complex $z$-plane are, respectively,
\begin{equation}
S_- = - \frac{\omega + \kappa B}{2 B} = - \frac{\tilde{\kappa} + \kappa}{2}, \qquad S_+ = \frac{\omega - \kappa B}{2 B} = \frac{\tilde{\kappa} - \kappa}{2}, \qquad S_\infty = \sqrt{\kappa^2 - \bar{m}^2 (r_0^2 + a^2)} = \mu.
\end{equation}
The phase integral formula can be given by~\cite{Kim2007}
\begin{equation} \label{productionrate}
\mathcal{N} = \mathrm{e}^{i S_\Gamma},
\end{equation}
where $S_\Gamma$ is the instanton action evaluated along the contour $\Gamma$ in the complex $z$-plane. By the two contour integrals shown in Fig.~\ref{fig1}, the phase integral formulae give
\begin{eqnarray}
\mathcal{N}_{1a} &=& \mathrm{e}^{i (-2 \pi i) (S_- + S_+ + S_\infty)} = \mathrm{e}^{-2 \pi (\kappa - \mu)} = \mathrm{e}^{-\frac{\bar{m}}{T_{\mathrm{KN}}}},
\\
\mathcal{N}_{1b} &=& \mathrm{e}^{i (-2 \pi i) (S_- + S_+ - S_\infty)} = \mathrm{e}^{-2 \pi (\kappa + \mu)} = \mathrm{e}^{-\frac{\bar{m}}{\bar T_{\mathrm{KN}}}},
\end{eqnarray}
which determine the Schwinger effect in the AdS space. Similarity, the two contour integrals shown in Fig.~\ref{fig2} give
\begin{eqnarray}
\mathcal{N}_{2a} &=& \mathrm{e}^{i (-2 \pi i) (S_- - S_+ - S_\infty)} = \mathrm{e}^{-2 \pi (\tilde \kappa + \mu)},
\\
\mathcal{N}_{2b} &=& \mathrm{e}^{i (-2 \pi i) (S_- - S_+ + S_\infty)} = \mathrm{e}^{-2 \pi (\tilde \kappa - \mu)},
\end{eqnarray}
which contribute to the Schwinger effect in the Rindler space.

\begin{figure}
\centering
\includegraphics[width=4in]{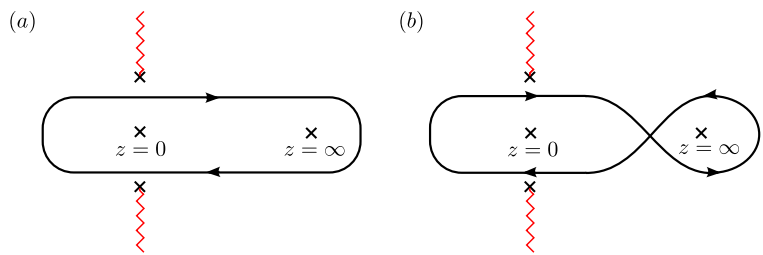}
\caption{The contour integral for the extremal KN black hole.}
\label{fig3}
\end{figure}

Few comments are in order. First, for the extremal KN black hole, the simple poles at inner/outer horizons degenerate into $z = 0$ for the extremal limit $B \to 0$ and the action becomes
\begin{equation}
S(z) = \oint \frac{dz}{z^2} \sqrt{(\omega - \kappa z)^2 - \bar{m}^2 (r_0^2 + a^2) z^2}.
\end{equation}
There are only two possible contour integrals shown in Fig.~\ref{fig3}, with $S_0 = - \kappa$, which give
\begin{eqnarray}
\mathcal{N}_{3a} &=& \mathrm{e}^{i (-2 \pi i)(S_0 + S_\infty)} = \mathrm{e}^{-2 \pi (\kappa - \mu)} = \mathrm{e}^{-\frac{\bar{m}}{T_\mathrm{KN}}},
\\
\mathcal{N}_{3b} &=& \mathrm{e}^{i (-2 \pi i)(S_0 - S_\infty)} = \mathrm{e}^{-2 \pi (\kappa + \mu)} = \mathrm{e}^{-\frac{\bar{m}}{\bar T_\mathrm{KN}}}.
\end{eqnarray}
Second, the limit of $a = 0$ gives a dyonic RN black hole. The emission of magnetic monopoles from an extremal magnetic black hole and the instability due to the vacuum polarization in the case of no magnetic monopoles was investigated in Ref.~\cite{Kim2004}. The production of monopole pairs in the Minkowski spacetime was originally studied by Affleck and Manton~\cite{Affleck:1981ag, Affleck:1981bma}. Third, the deep connection between the phase integral method for particle production~\cite{Kim2007, Kim2013} and the black hole monodromy~\cite{Castro2013} remains an open question to pursue further, which is beyond the scope of this paper. Finally, a passing remark is that within the framework of QED the effect of a strong magnetic field on the Schwinger effect in a uniform electric field in a de Sitter space is both the suppression of mean number of pairs analogous to the transverse momentum in Minkowski spacetime and the enhancement due to the density of states proportional to the magnetic field~\cite{Bavarsad2017}. Considering the duality of pair production and vacuum polarization under the scalar curvature~\cite{Cai2014}, we may expect a similar enhancement of pair production due to a magnetic field in the AdS space, which requires a further study. In fact, the Schwinger effect is enhanced by the black hole's magnetic charge due to the increased Unruh temperature in the effective temperature~(\ref{effectivetemperature}).

\section{Dual CFTs Descriptions}\label{sec5}
Now we will calculate the the mean number of produced pairs and the absorption cross-section ratio of the scalar operators from the dual CFT side via the KN/CFTs duality~\cite{Chen:2010ywa, Chen:2012np}, in which the dual CFTs descriptions are called the angular momentum $J$-picture in terms of the Kerr/CFT correspondence~\cite{Guica:2008mu, Compere:2012jk} and the electric charge $Q$-picture realized in the RN/CFT correspondence~\cite{Chen:2009ht, Chen:2011gz}. Recall that for the dyonic RN black hole, there is an additional dual CFT description associated with the magnetic charge, i.e. the $P$-picture, which is related to the $Q$-picture by the electromagnetic duality in four dimensional spacetime~\cite{Chen:2010yu, Chen:2012pt}. Therefore, for the dyonic KN black hole, it should also have threefold dual CFTs descriptions.

From the field/operator duality in the AdS/CFT correspondence, the scalar field in the near extremal KN black hole is dual to a scalar operator in the two-dimensional CFT with conformal dimensions $(h_\mathrm{L}, h_\mathrm{R})$ which can be determined from the asymptotic expansion of the bulk dyonic charged scalar field at the AdS boundary. While for the pair production process, the conformal dimensions become complex and have the same magnitudes, so that there are two choices
\begin{equation}\label{cfmdim}
h_\mathrm{L} = h_\mathrm{R} = \frac12 \pm i \mu,
\end{equation}
we choose the plus sign without loss of generality and the case of the minus sign will be mentioned at the end of this section.

Note that the absorption cross section ratio of the dual scalar operator in the two-dimensional CFT has the universal form, i.e.
\begin{equation} \label{CFTabs}
\sigma_\mathrm{abs} \sim T_\mathrm{L}^{2 h_\mathrm{L} - 1} T_\mathrm{R}^{2 h_\mathrm{R} - 1} \sinh\left( \frac{{\tilde\omega}_\mathrm{L}}{2 T_\mathrm{L}} + \frac{{\tilde\omega}_\mathrm{R}}{2 T_\mathrm{R}} \right) \left| \Gamma\left( h_\mathrm{L} + i \frac{{\tilde\omega}_\mathrm{L}}{2 \pi T_\mathrm{L}} \right) \right|^2 \left| \Gamma\left( h_\mathrm{R} + i \frac{{\tilde\omega}_\mathrm{R}}{2 \pi T_\mathrm{R}} \right) \right|^2,
\end{equation}
where $T_\mathrm{L}$ and $T_\mathrm{R}$ are left- and right-hand temperature of the dual CFT, ${\tilde\omega}_\mathrm{L} = \omega_\mathrm{L} - q_\mathrm{L} \Phi_\mathrm{L}$ and ${\tilde\omega}_\mathrm{R} = \omega_\mathrm{R} - q_\mathrm{R} \Phi_\mathrm{R}$ are the total excited energy of the left- and right-hand sectors in which $(q_\mathrm{L}, q_\mathrm{R})$ and $(\Phi_\mathrm{L}, \Phi_\mathrm{R})$ are respectively the charges and chemical potential (both including the electric and the magnetic ones for the dyonic KN black hole case) of the dual left and right-hand operators.

To compare Eq.~(\ref{CFTabs}) with Eq.~(\ref{absorption}), the thermodynamic properties of the near extremal KN black hole are needed: the black hole entropy and temperature\footnote{Note that a direct calculation shows $S_\mathrm{BH} = \pi (\hat{r}_+^2 + a^2) \sim \pi (r_0^2 + a^2 + 2 r_0 \varepsilon B)$. However, this entropy is associated with the time coordinate $\hat{t}$ in the original dyonic KN black hole Eq.~(\ref{dyonic KN}), while for the near horizon near extremal KN geometry, its entropy and the first law of thermodynamics are associated with the time coordinate $t$ in Eq.~(\ref{NHKN}). Therefore, $S_\mathrm{BH}$ and $T_\mathrm{H}$ should have the value in Eq.~(\ref{KNentropy}).}
\begin{eqnarray}\label{KNentropy}
&& S_\mathrm{BH} = \pi (\hat{r}_+^2 + a^2) \quad \Rightarrow \quad S_\mathrm{BH} \sim \pi (r_0^2 + a^2 + 2 r_0 B),
\\
&& T_\mathrm{H} = \frac{\hat{r}_+ - \hat{r}_-}{4 \pi (\hat{r}_+^2 + a^2)} \quad \Rightarrow \quad T_\mathrm{H} = \frac{B}{2 \pi}.
\end{eqnarray}
Next, we will analyze the dual CFTs descriptions of the pair production rate and the absorption cross section ratios in the near extremal dyonic KN black hole in the $J$- and $Q$- as well as the $P$-pictures, respectively.

\subsection{$J$-picture}
In the $J$-picture, the left- and right-hand central charges of the dual CFT are~\cite{Chen:2010ywa, Chen:2012np}
\begin{equation}
c_\mathrm{L}^J = c_\mathrm{R}^J = 12 J,
\end{equation}
and the associated left- and right-hand temperatures are
\begin{equation}
T_\mathrm{L}^J = \frac{\hat{r}_+^2 + \hat{r}_-^2 + 2 a^2}{4 \pi a (\hat{r}_+ + \hat{r}_-)}, \qquad T_\mathrm{R}^J = \frac{\hat{r}_+ - \hat{r}_-}{4 \pi a}.
\end{equation}
For the near horizon geometry of the near extremal dyonic KN black hole, the results are
\begin{equation}
T_\mathrm{L}^J = \frac{r_0^2 + a^2}{4 \pi a r_0}, \qquad T_\mathrm{R}^J = \frac{B}{2 \pi a}.
\end{equation}
Then from the Cardy formula, the CFT microscopic entropy is
\begin{equation}
S_\mathrm{CFT} = \frac{\pi^2}3 (c_\mathrm{L}^J T_\mathrm{L}^J + c_\mathrm{R}^J T_\mathrm{R}^J) = \pi (r_0^2 + a^2 + 2 r_0 B),
\end{equation}
which matches with the macroscopic entropy~(\ref{KNentropy}) of the near extremal KN black hole.

Besides, the variation of the entropy in both sides should also agree with each other according to the duality, namely, the first law of thermodynamics of the black hole equals that of the dual CFT, i.e., $\delta S_\mathrm{BH} = \delta S_\mathrm{CFT}$
\begin{equation}\label{1stlaws}
\frac{\delta M - \Omega_\mathrm{H} \delta J - \Phi_\mathrm{H} \delta Q - \bar\Phi_\mathrm{H} \delta P}{T_\mathrm{H}} = \frac{\tilde{\omega}_\mathrm{L}}{T_\mathrm{L}} + \frac{\tilde{\omega}_\mathrm{R}}{T_\mathrm{R}},
\end{equation}
where $\tilde{\omega}_\mathrm{L}$ and $\tilde{\omega}_\mathrm{R}$ are the excitation of the total energy in the left- and right-hand sectors of the dual CFT, and the angular velocity and chemical potentials at $r = B$ are respectively
\begin{equation}
\Omega_\mathrm{H} = - \frac{2 a r_0 B}{r_0^2 + a^2}, \qquad \Phi_\mathrm{H} = \frac{Q (Q^2 + P^2) B}{r_0^2 + a^2}, \qquad \bar\Phi_\mathrm{H} = \frac{P (Q^2 + P^2) B}{r_0^2 + a^2}.
\end{equation}
In the $J$-picture, we have $T_\mathrm{L} = T_\mathrm{L}^J$ and $T_\mathrm{R} = T_\mathrm{R}^J$. The trick to probe the rotation only is to turn off the charges of the probe scalar field, then the variation of the conserved charges of the dyonic KN black hole are $\delta M = \omega, \; \delta J = n, \; \delta Q = 0, \; \delta P = 0$. Subsequently, one has
\begin{eqnarray}
&&{\tilde \omega}_\mathrm{L}^J = n \quad \mathrm{and} \quad {\tilde \omega}_\mathrm{R}^J = \frac{\omega}{a},
\\
&&\frac{{\tilde\omega}_\mathrm{L}^J}{2 T_\mathrm{L}^J} = - \pi \kappa \quad \mathrm{and} \quad \frac{{\tilde\omega}_\mathrm{R}^J}{2 T_\mathrm{R}^J} = \pi \tilde\kappa,
\end{eqnarray}
where $q, p$ are set to zero. Consequently, the absorption cross section ratio~(\ref{absorption}) of the scalar field (with $q = p = 0$) in the near extremal dyonic KN black hole matches well with that of its dual scalar operator in Eq.~(\ref{CFTabs}) in the $J$-picture.

\subsection{$Q$-picture}
In the $Q$-picture, the left- and right-hand central charges of the dual CFT are~\cite{Chen:2010ywa, Chen:2012np, Chen:2010yu}
\begin{equation}
c_\mathrm{L}^Q = c_\mathrm{R}^Q = \frac{6 Q (Q^2 + P^2)}{\ell},
\end{equation}
where the parameter $\ell$ is the measure of the $U(1)$ bundle formed by the background Maxwell field, which can be interpreted as the radius of the embedded extra circle in the fifth dimension. The associated temperatures of the left- and right-hand sectors are
\begin{equation}
T_\mathrm{L}^Q = \frac{(\hat{r}_+^2 + \hat{r}_-^2 + 2 a^2) \ell}{4 \pi Q (\hat{r}_+ \hat{r}_- - a^2)}, \qquad T_\mathrm{R}^Q = \frac{(\hat{r}_+^2 - \hat{r}_-^2) \ell}{4 \pi Q (\hat{r}_+ \hat{r}_- - a^2)}.
\end{equation}
For near horizon geometry of the near extremal dyonic KN black hole, they are
\begin{equation}
T_\mathrm{L}^Q = \frac{(r_0^2 + a^2) \ell}{2 \pi Q (Q^2 + P^2)}, \qquad T_\mathrm{R}^Q = \frac{r_0 B \ell}{\pi Q (Q^2 + P^2)}.
\end{equation}
Then the microscopic entropy of the dual CFT computed from the Cardy formula
\begin{equation}
S_\mathrm{CFT} = \frac{\pi^2}3 (c_\mathrm{L}^Q T_\mathrm{L}^Q + c_\mathrm{R}^Q T_\mathrm{R}^Q) = \pi (r_0^2 + a^2 + 2 r_0 B),
\end{equation}
again reproduces the area entropy of the near extremal dyonic KN black hole.

Now in the $Q$-picture, the modes $n$ and $p$, which characterize the rotation and magnetic charge of the charged probe scalar field, should be turned off in order that the electrically charged probe only detect the electric charge of the black hole, i.e., $\delta M = \omega, \; \delta J = 0, \; \delta Q = q, \; \delta P = 0$. Then, using Eq.~(\ref{1stlaws}), with $T_\mathrm{L} = T_\mathrm{L}^Q$ and $T_\mathrm{R} = T_\mathrm{R}^Q$, the excitations of the total energy of the dual $Q$-picture CFT are
\begin{equation}
{\tilde \omega}_\mathrm{L}^Q = - q \ell \quad \mathrm{and} \quad {\tilde \omega}_\mathrm{R}^Q = \frac{2 \omega r_0 \ell}{Q (Q^2 + P^2)},
\end{equation}
namely, $\frac{{\tilde\omega}_\mathrm{L}^Q}{2 T_\mathrm{L}^Q} = - \pi \kappa$ and $\frac{{\tilde\omega}_\mathrm{R}^Q}{2 T_\mathrm{R}^Q} = \pi \tilde\kappa$. Therefore, the absorption cross section ratio of the charged scalar field (with $n = p = 0$) in Eq.~(\ref{absorption}) also matches well with that of its dual scalar operator in Eq.~(\ref{CFTabs}) in the $Q$-picture.

\subsection{$P$-picture}
For the $P$-picture CFT, the left- and right-hand central charges can be obtained from the electromagnetic duality, which are
\begin{equation}
c_\mathrm{L}^P = c_\mathrm{R}^P = \frac{6 P (Q^2 + P^2)}{\ell},
\end{equation}
and the associated left- and right-hand temperatures are
\begin{equation}
T_\mathrm{L}^P = \frac{(\hat{r}_+^2 + \hat{r}_-^2 + 2 a^2) \ell}{4 \pi P (\hat{r}_+ \hat{r}_- - a^2)}, \qquad T_\mathrm{R}^P = \frac{(\hat{r}_+^2 - \hat{r}_-^2) \ell}{4 \pi P (\hat{r}_+ \hat{r}_- - a^2)},
\end{equation}
which, in the near extremal case, can be expressed as
\begin{equation}
T_\mathrm{L}^P = \frac{(r_0^2 + a^2) \ell}{2 \pi P (Q^2 + P^2)}, \qquad T_\mathrm{R}^P = \frac{r_0 B \ell}{\pi P (Q^2 + P^2)}.
\end{equation}
Then we have the matching between the microscopic and macroscopic entropies
\begin{equation}
S_\mathrm{CFT} = \frac{\pi^2}3 (c_\mathrm{L}^P T_\mathrm{L}^P + c_\mathrm{R}^P T_\mathrm{R}^P) = \pi (r_0^2 + a^2 + 2 r_0 B).
\end{equation}

In the $P$-picture, similarly to the $Q$-picture, we should turn off the modes of $n$ and $q$, i.e. considering $\delta M = \omega, \; \delta J = 0, \; \delta Q = 0, \; \delta P = p$. Then, applying Eq.~(\ref{1stlaws}), with $T_\mathrm{L} = T_\mathrm{L}^P$ and $T_\mathrm{R} = T_\mathrm{R}^P$, the excitations of the total energy of the dual $P$-picture CFT are
\begin{equation}
{\tilde \omega}_\mathrm{L}^P = - p \ell \quad \mathrm{and} \quad {\tilde \omega}_\mathrm{R}^P = \frac{2 \omega r_0 \ell}{P (Q^2 + P^2)},
\end{equation}
namely, $\frac{{\tilde\omega}_\mathrm{L}^P}{2 T_\mathrm{L}^P} = - \pi \kappa$ and $\frac{{\tilde\omega}_\mathrm{R}^P}{2 T_\mathrm{R}^P} = \pi \tilde\kappa$. Therefore, in the $P$-picture of the KN/CFTs duality, the absorption cross section ratio of the charged scalar field (with $n = q = 0$) in Eq.~(\ref{absorption}) also matches well with that of its dual scalar operator in Eq.~(\ref{CFTabs}).

Moreover, using the relation between pair production rate and the absorption cross section ratio, i.e., $|\beta|^2(\mu \to -\mu) \to -\sigma_\mathrm{abs}$, the holographic descriptions of $|\beta|^2$ can be understood in all of the $J$-, $Q$- and $P$-pictures. To have a more clear picture of pair production rate from the dual CFT side, it is interesting to notice that changing $\mu \to -\mu$ just corresponds to changing the conformal dimension of the dual operators
\begin{equation}
h_\mathrm{L} = h_\mathrm{R} = \frac12 + i \mu \quad \to \quad h_\mathrm{L} = h_\mathrm{R} = \frac12 - i \mu,
\end{equation}
which indicates that from the dual CFT side, the pair production rate from the operators with $h_\mathrm{L} = h_\mathrm{R} = \frac12 + i \mu$ equals the (absolute value) of the absorption cross ratio from the operators with $h_\mathrm{L} = h_\mathrm{R} = \frac12 - i \mu$, i.e., those operators are related by adopting different quantization conditions, the standard quantization and the alternative quantization in the AdS/CFT correspondence~\cite{Faulkner:2010jy}.

\section{Conclusion}\label{sec6}
We investigated the Schwinger pair production of the dyonic charged scalar pairs in the near horizon region of the near extremal dyonic KN black hole. It generalized the previous study in the near extremal KN black hole~\cite{Chen:2016caa} to the background geometry containing both the electric and magnetic charges and the probe scalar field also carrying the magnetic charge. We found that the violation of BF bound is still equivalent to the condition to ensure that the produced dyonic charged particle pairs could have propagating modes for particle and antiparticle states~\cite{Chen:2016caa}. The physical quantities, i.e. the vacuum persistence amplitude, mean number of produced pairs, and absorption cross section ratio are evaluated, which have the same forms as those of the KN black holes without magnetic charges. However, the parameters that encode the gravity information have been deformed by the magnetic charges.

The thermal interpretation of the pair production rate for the dyonic KN black hole can also be proposed by appropriately separating instanton actions into two parts; the one is the Schwinger effect in the AdS$_2$ space caused mainly by the electromagnetic field of black holes and the other is the Schwinger effect in the Rindler space due to the Hawking temperature, i.e. the deviation from the extremality. The leading term of the production rate is determined by the effective temperature $T_\mathrm{KN}$, which is a kind of Unruh temperature originated from the acceleration due to the electromagnetic force and the AdS$_2$ curvature. As expected, the magnetic field affects on the magnetic charged particles in the same way as the electric field on the electric charged particles. In fact, there is duality between the electric charges and magnetic charges of the hole and produced pairs. The effect of the AdS$_2$ curvature is to bind pairs and decrease the pair production while the electromagnetic field increases the Unruh temperature for the effective temperature and thus enhances the pair production. The angular momentum, electric charge, and magnetic charge of the produced dyonic particles are expected to couple with the corresponding thermal quantities, i.e. the angular velocity, electric chemical potential, and magnetic chemical potential, of the dyonic KN black hole as the subleading terms to the leading Boltzmann factor.

The twofold dual CFTs descriptions (i.e., $J$-picture and $Q$-picture) of the Schwinger pair production of the near extremal KN black hole~\cite{Chen:2016caa} are extended to the near extremal dyonic KN black hole which includes an additional magnetic hair and consequently has threefold dual CFTs descriptions. The presence of the $P$-picture is associated with the dual gauge potential which can be regarded as a new ``hair'' of the dyonic KN black hole and provides the $U(1)$ fiber on the base manifold; in particular, it can only be probed by the magnetic charge of the scalar field. Based on the threefold dyonic KN/CFTs duality, the absorption cross section ratios and the pair production rate of the dyonic charged scalar field and its dual operators in the CFT are matched well with each other in the $J$-, $Q$-, and $P$-pictures, respectively.

\section*{Acknowledgments}
The authors would like to thank Yong-Ming Cho for suggestion to consider the monopole production.
The work of C.M.C. was supported by the Ministry of Science and Technology of Taiwan under the grant MOST 105-2112-M-008-018. The works of S.P.K. was supported by the Basic Science Research Program through the National Research Foundation of Korea (NRF) funded by the Ministry of Education (NRF-2015R1D1A1A01060626). J.R.S. was supported by the National Natural Science Foundation of China (No.~11675272), the Open Project Program of State Key Laboratory of Theoretical Physics, Institute of Theoretical Physics, Chinese Academy of Sciences, China (No.~Y5KF161CJ1) and the Fundamental Research Funds for the Central Universities.


\end{document}